\documentclass[sigconf,natbib=true,anonymous=false]{acmart}

\usepackage{graphicx}
\usepackage{soul,xcolor}
\usepackage{balance}

\usepackage[noadjust]{marginnote}

\newcommand{\g}[1]{\textcolor{black}{#1}}
\newcommand{\jm}[1]{\textcolor{black}{#1}}
\newcommand{\jmedit}[1]{\textcolor{black}{#1}}

\newcommand{\release}[1]{\textcolor{black}{#1}}

\AtBeginDocument{%
  \providecommand\BibTeX{{%
    \normalfont B\kern-0.5em{\scshape i\kern-0.25em b}\kern-0.8em\TeX}}}

\setcopyright{acmcopyright}
\copyrightyear{2023}
\acmYear{2023}
\acmDOI{XXXXXXX.XXXXXXX}

\begin{document}
\setstcolor{red}

\title{SARA: A Collection of Sensitivity-Aware Relevance Assessments}

\author{Jack McKechnie}
\email{j.mckechnie.1@research.gla.ac.uk}
\orcid{1234-5678-9012}
\affiliation{%
  \institution{University of Glasgow}
  \city{Glasgow}
  \state{Scotland}
  \country{UK}
}

\author{Graham McDonald}
\email{graham.mcdonald@glasgow.ac.uk}
\orcid{0000-0002-1266-5996}
\affiliation{%
  \institution{University of Glasgow}
  \city{Glasgow}
  \state{Scotland}
  \country{UK}
}

\renewcommand{\shortauthors}{McKechnie and McDonald}

\begin{abstract}
Large archival collections, such as email or government documents, must be manually reviewed to identify any sensitive information before the collection can be released publicly. Sensitivity classification has received a lot of attention in the literature. However, more recently, there has been increasing interest in developing sensitivity-aware search engines that can provide users with relevant search results, while ensuring that no sensitive documents are returned to the user. Sensitivity-aware search would mitigate the need for a manual sensitivity review prior to collections being made available publicly. To develop such systems, there is a need for test collections that contain relevance assessments for a set of information needs as well as ground-truth labels for a variety of sensitivity categories. The well-known Enron email collection contains a classification ground-truth that can be used to represent sensitive information, e.g., the \textit{Purely Personal} and \textit{Personal but in Professional Context} categories can be used to represent sensitive personal information. However, the existing Enron collection does not contain a set of information needs and relevance assessments. In this work, we present a collection of fifty information needs (topics) with crowdsourced query formulations (3 per topic) and relevance assessments (11,471 in total) for the Enron collection (mean number of relevant documents per topic = 11, $\sigma^2$ = 34.7). The developed information needs, queries and relevance judgements are available on GitHub and will be available along with the existing Enron collection through the popular ir\_datasets library. Our proposed collection results in the first freely available test collection for developing sensitivity-aware search systems.
\end{abstract}

\ccsdesc[500]{Information systems~Test collections}

\keywords{test collection, relevance, sensitivity}

\maketitle

\section{Introduction}\label{sec:introduction}
Sensitive information, for example information about a person's medical history, finances or citizenship, is often present in large collections of documents\g{,} such as email archives, \jm{internal company \g{or government} documents\g{, or documents that are requested through e-discovery~\cite{oard2010evaluation,e-discovery} in litigation trials}.} \jm{Currently, collections that potentially contain sensitive information cannot be made available publicly without first being manually reviewed by experts to ensure that no sensitive documents are released.} \g{Sensitivity review} is a time-consuming and expensive process \g{that can} prohibit many such collections from ever being released. For example, the 2014 US Presidential candidate Hilary Clinton released a collection of emails from when she was working for the government. Clinton sent 30,490 emails to the US State Department to be reviewed and released as quickly as possible. However, the reviewing process took nearly a full year with 25 people working on the review. 

Archival document collections can be a valuable resource if they are made available to be searched. For example, email collections can be valuable to historians as they serve as a permanent record of an organisation's or person’s activities~\cite{gollins_archive} and medical records can be useful for medical research if they are made available~\cite{medical_collections}. Indeed, the UK Government has noted that some of the data that would be of the most value cannot be released due to concerns about sensitivity~\cite{gov_ai_deal}. Most of the previous work on identifying sensitive information has focused on sensitivity classification. Sensitivity classification has been shown to be useful for helping sensitivity reviewers to increase the speed and accuracy of their review~\cite{mcdonald_zero_tolerance,narvala_categories} and to \release{identify} which documents should be prioritised for review to increase the number of documents that can be released with finite reviewing resources~\cite{narvala_prioritising}. However, recently, there has been an increasing interest in developing search systems that are able to index entire collections (including any potential sensitivities) and deploy a retrieval model that is capable of providing the search engine users with relevant results while ensuring that no sensitive documents are included in the search results, for example in~\cite{oard_sas}. We refer to this as sensitivity-aware search.  

The development of sensitivity-aware search engines requires test collections that contain relevance assessments for a set of information needs along with ground-truth labels for categories of sensitivities. There are a couple of test collections available that have made progress towards this goal. However, each of these collections has disadvantages for some researchers. The TREC 2010 Legal Track provides a collection of documents labelled for both relevance and for legal privilege \cite{trec_legal_2010}. Legal privilege is a form of sensitive information. However, it is very specific to the e-discovery context and, therefore, the collection is not suitable for general sensitivity tasks. Moreover, since the collection was assessed separately for relevance and for sensitivity, only a small subset of the documents have both relevance and sensitivity assessments. The Avocado Research Email Collection~\cite{avocado_collection,oard_collection} contains relevance and sensitivity assessments. However, the Avocado collection is somewhat costly to obtain, which makes its use prohibitive to researchers that are not in a position to purchase it. Moreover, the collection is restricted in what it can be used for due to its licence agreement (for example, the collection cannot be used for crowdsourced experiments).

Another potentially useful test collection is the UC Berkeley version of the Enron email collection.\footnote{\url{https://bailando.berkeley.edu/enron_email.html}} The UC Berkley Enron collection provides classification labels for a rich taxonomy of classification categories~\cite{hearst_teaching}, some of which are representative of sensitive information. For example, the \textit{Purely Personal} and \textit{Personal but in Professional Context} categories are representative of sensitive personal information. However, the existing collection does not contain a set of information needs and relevance assessments that are necessary for developing sensitivity-aware search systems.

In this work, we present an extension to the UC Berkeley version of the Enron email test collection\footnote{There were a number of prior releases of the Enron Collection. The May 7 2015 version is the stable version of the collection and earlier versions of the collection should not be used for research. The UC Berkeley version of the Enron collection is a subset of the May 7 2015 collection. For the sake of brevity, in this work, we refer to the UC Berkeley version of the collection simply as the labelled Enron collection.} to make the collection suitable for the development of sensitivity-aware search. Using a topic modelling approach, we identify fifty topics of discussion in the Enron emails and manually create short passages of text to represent each of the identified information needs (topics). Three representative query formulations are crowdsourced for each of the topics and a pooling approach is used to obtain crowdsourced relevance judgements (11,471 in total). The mean number of relevant documents per topic = 11 ($\sigma^2$ = 34.7). Our extension to the Enron collection provides sensitivity-aware relevance assessments (SARA) that make the labelled Enron collection a valuable resource for the development of sensitivity-aware search. 


The remainder of this paper is structured as follows: In Section~\ref{sec:RelatedWork}, we discuss related work on sensitivity classification and sensitivity-aware search. In Section~\ref{sec:EnronCollection} we present the existing Enron Email collection before, in Section~\ref{sec:SASCollection}, presenting the process for constructing our extension to the collection for sensitivity-aware search. In Section~\ref{sec:SASCollectionAnalysis}, we present some analysis of our extension, before providing some additional discussions in Section~\ref{sec:Discussion} and concluding words in Section~\ref{sec:Conclusions}.


\section{Related Work/Background}\label{sec:RelatedWork}
In this section, we first discuss previous work relating to sensitivity classification before, secondly, discussing work relating to sensitivity-aware search.

\citet{gollins_archive} identified some of the challenges that are associated with sensitivity reviewing archival collections of digital documents, and how automatic sensitivity classification approaches may be able to assist with the human review process.

Since then there have been numerous works that have investigated using sensitivity classifiers to assist sensitivity reviewers, for example by highlighting documents which are likely to need to be reviewed and by automatically classifying documents entirely. For example,~\citet{mcdonald_towards,mcdonald_ngrams,mcdonald_semantic_features} proposed several sensitivity classification approaches in the context of predicting if information is exempt from being publicly released through the United Kingdom Freedom of Information Act 2000\footnote{https://www.legislation.gov.uk/ukpga/2000/36/part/II}. \citet{mcdonald_towards} proposed using classification features which take into account the expected sensitivity-related risk that was related to the countries which were mentioned in the documents' text. Syntactic and semantic document features have also been shown to be useful for improving sensitivity classification~\cite{mcdonald_ngrams,mcdonald_semantic_features}. 

\citet{narvala_categories} investigated the identification of latent semantic categories in document collections to improve the efficiency of human sensitivity reviewers and \citet{mcdonald_active_learning} developed active learning strategies to reduce the amount of reviewing effort that is required to be able to train an effective sensitivity classifier. All of these works (\cite{mcdonald_towards,mcdonald_ngrams,mcdonald_semantic_features,narvala_categories,mcdonald_active_learning}) used a test collection of 3801 real government documents that contained real government sensitivities. While it is important for these works to be able to evaluate their proposed approaches on real-life sensitivities, the highly sensitive content in the collection means that the collection is not easily sharable among researchers. Moreover, although the collection has a sensitivity classification ground truth, it does not contain a ground truth of information needs with relevance assessments, and is therefore not suitable for developing sensitivity-aware search approaches.

There has been comparatively little previous work investigating sensitivity-aware search approaches. \citet{oard_sas} proposed a variant of the Discounted Cumulative Gain metric~\cite{ndcg}, named Cost Sensitive Discounted Cumulative Gain ($CS\text{-}nDCG$), that incorporated both the relevance of the documents in a ranking as well as the preponderance of sensitive document that are in the ranking. The $CS\text{-}nDCG$ metric facilitates the evaluation of retrieval models that try to filter out sensitive documents from the list of search results. \citet{oard_sas} also proposed a novel Learning-to-Rank approach that considered the probability of a document being sensitive, as defined by a pre-trained classifier, and integrated a loss function that optimised for normalised $CS\text{-}nDCG$. However, unable to use a collection containing real sensitivities, \citet{oard_sas} evaluated their proposed approach using the OHSUMED dataset and a subset of PubMed labels (categories of diseases) as a proxy for sensitive categories. This highlighted the lack of, and need for, a freely available and easily accessible test collection that contains sensitivity labels and relevance assessments for a set of queries, to evaluate sensitivity-aware search. 

To try to address the lack of an available test collection for sensitivity-aware search, \citet{oard_collection} developed an extension to the Avocado Research Email Collection\footnote{\url{https://catalog.ldc.upenn.edu/LDC2015T03}} that included judgments for sensitivity and relevance assessments for a set of information needs. In \cite{oard_collection}, the sensitivity labels were based on two fictional personas that were developed to take into account the behaviour patterns of people with a range of levels of risk-aversion as to what they consider to be sensitive. The Avocado collection from~\citet{oard_collection} provides a useful test collection for evaluating sensitivity-aware search. However, it is somewhat costly to obtain, which makes its use prohibitive to researchers in certain cases, and the collection is also restricted in what it can be used for due to its licensing arrangements. For example, the collection cannot be used in crowdsourced experiments.

Finally, it is worth noting that, although~\citet{oard_collection} introduced a collection that is suitable for sensitivity-aware search, our sensitivity-aware relevance assessments extension to the Enron collection provides a valuable new resource for the field for two main reasons: (1) our sensitivity-aware relevance assessments and the Enron email collection are both made available to the community free of charge and without any prohibitively restrictive licensing arrangements. This makes our extended Enron collection suitable as a first test collection of choice for exploratory researchers that are wanting to evaluate their ideas without having to commit to a substantive financial outlay to do so, or are wanting to perform crowdsourced evaluations of their experiments, and (2) sensitivity classifiers that are developed and evaluated on a narrowly defined set of sensitivities are unlikely to generalise well since sensitivity is often very broadly defined and can be subjective. The Enron text collection provides a range of different classification categories that are representative of different types of sensitivities. This provides opportunities to deploy experimental setups that explicitly evaluate how generalisable, in terms of sensitivity, the developed models are likely to be. Moreover, researchers can evaluate this further by using our extended version of the Enron collection alongside the Avocado collection from~\citet{oard_collection}.


\section{Existing Enron Email Collection}\label{sec:EnronCollection}

\begin{figure}[tb]
    \centering
    \includegraphics[width=0.5\textwidth]{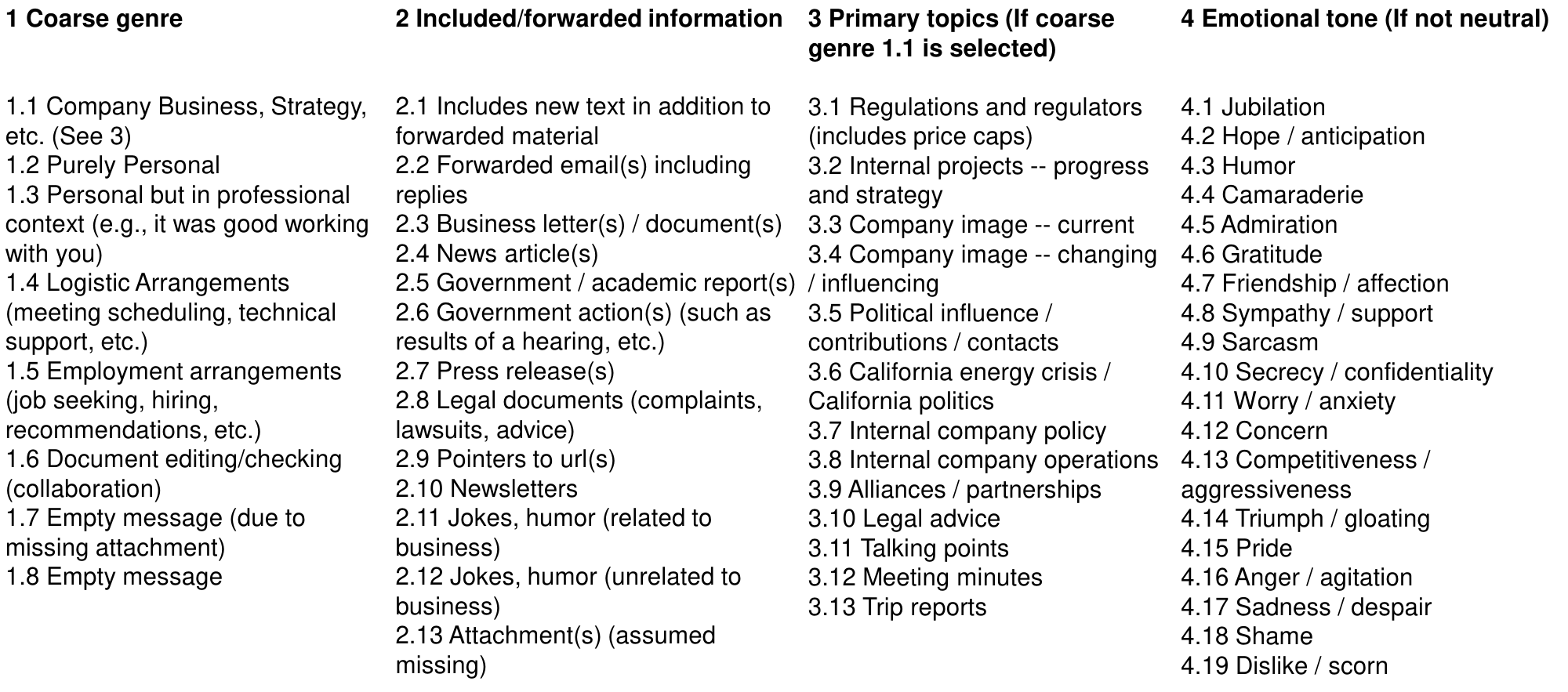}
    \caption{Enron Classification Categories.}
    \label{fig:enroncats}
\end{figure}

The original version of the Enron Email Collection was made available by the Federal Energy Regulatory Commission\footnote{https://www.ferc.gov/} (FERC) as a public archive of the commission's investigation into the Western Energy Crisis of 2000/2001~\cite{ferc_report}. Originally, the collection \g{contained} $\sim$1.5 million emails. However, there were also many duplicate emails and unused folder structures. An updated version of the Enron email collection with approximately 500,000 emails was made publicly available after the collection was acquired by researchers at the Massachusetts Institute of Technology\footnote{https://www.mit.edu/} (MIT)~\cite{introducing_enron}. The updates to the MIT version of the collection included converting invalid email addresses to valid placeholders, removing any attached files and removing a number of emails that Enron employees had requested to be removed. Subsequently, several updated versions of the collection were released with further filtering of the emails that were contained in the collection. The current version of the Enron email collection~\cite{introducing_enron} was released in 2015 and is available from the Carnegie Mellon University (CMU) website.\footnote{\url{http://www.cs.cmu.edu/~enron/}} 

In order to aid the development of sensitivity-aware search engines, we aim to create a test collection which includes relevance assessments for topics, as well as sensitivity labels. For this to be possible, a collection with a number of different characteristics is needed. A variety of classes which can reasonably be considered as sensitive are essential as the collection needs to provide labels for sensitivity which reflect true sensitive information. The \citet{hearst_teaching} labelled version of the Enron Email Collection is a subset \jm{of the CMU collection} that contains 1702 emails that were annotated as part of a class project at UC Berkley. Students in the Natural Language Processing course were tasked with annotating the emails as relevant or not relevant to 53 different categories. Therefore, the~\citet{hearst_teaching} labelled version of the Enron email collection provides a rich taxonomy of labels which can be used for multiple definitions of sensitivity such as the \textit{Purely Personal} and \textit{Personal but in a Professional Context}. Figure~\ref{fig:enroncats} presents the categories that the emails were annotated for. 

The collection also needs to have a diverse enough set of topics that are discussed in the documents, in order to develop information needs based on the topics. Previous works have identified plentiful and rich topics in the collection using topic modelling~\cite{original_enron_collection_stats}, further motivating the choice of the labelled Enron collection. The context of the creation of the emails in this collection also influences our decision to extend it. The emails were created in a real-world context, with employees in an actual company interacting with each other. This is reflective of personal and corporate email collections which are currently not available to be released due to their sensitive information. Finally, the collection must not be held under a restrictive license and must be available for little to no cost so that other researchers are able to use the collection to develop sensitivity-aware search solutions. The labelled Enron collection is freely available and not held under a restricting licence. As can be seen from Figure~\ref{fig:enroncats}, there are \jm{three high-level categories in the collection: \textit{Coarse Genre}, \textit{Included/forwarded information}, and \textit{Emotional Tone}. Each of the high-level categories has an associated set of up to nineteen subcategories that an email can be associated with.\footnote{The subcategories of \textit{Primary topics} are only applied to emails that are labelled as the subcategory 1.1 \textit{Company Business, Strategy}, and the \textit{Emotional Tone} subcategories are only applied to emails that are not neutral in tone.} The categories and their subcategories were hand crafted by~\citet{hearst_teaching} and refined after discussions with the class. Each email was read and annotated by at least two students.}


\begin{figure}[tb]
    \centering
    \includegraphics[width=0.5\textwidth]{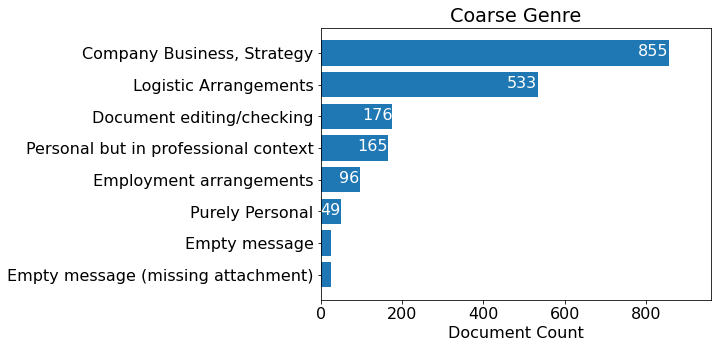}
\caption{Distribution of classification labels across the subcategories of the Coarse Genre category}
\label{fig:course_genre}
\end{figure}

When using the labelled Enron collection as a collection for identifying sensitive information, there are a number of the subcategories labels that could reasonably be chosen to represent sensitive categories. \jm{For example, the \textit{Shame} and \textit{Worry / Anxiety} subcategories of the \textit{Emotional Tone} category. However, 1699 of the emails are categorised for their course genre (e.g., \textit{Company Business, Strategy, etc}, \textit{\textit{Logistic Arrangement}} or \textit{Purely Personal}), whereas only 310 emails are annotated for their emotional tone.} \release{Therefore, in this work, we consider the \textit{Coarse Genre} subcategories of \textit{Purely Personal} and \textit{Personal but in a Professional Context} to be sensitivity labels and emails with these labels are referred to as sensitive emails. The full distribution of classification labels of the \textit{Coarse Genre} categories can be seen in Figure~\ref{fig:course_genre}}. There are 211 emails that \jm{have the \textit{Purely Personal} and \textit{Personal but in a Professional Context} labels}, more than other potentially sensitive categories, and because personal sensitivities have been researched in the literature significantly, e.g., in~\cite{mcdonald_towards, mcdonald_semantic_features, pii}. The \textit{Purely Personal} category includes invitations to weddings and events and conversations between family members. These are emails that are personal but do not include any relation to work being done at Enron. Conversely, the \textit{Personal but in a Professional Context} category contains emails which are personal, but are related to work that was being done at Enron. This includes comments about the quality of people’s work and expressions of feelings about employee treatment. 

\section{Sensitivity-Aware Search Test Collection}\label{sec:SASCollection}
In this section, we present our sensitivity-aware relevance assessments (SARA) extension to the labelled Enron email collection. We deploy a topic modelling approach to identify topical themes in the labelled Enron collection that serve as a basis for our information needs and relevance assessments. Two separate crowdsourcing tasks are carried out in the development of SARA. Firstly, query formulations are crowdsourced to represent the information needs and, secondly, relevance assessments are crowdsourced for a pooled set of documents from the labelled Enron collection for each of the information needs. We discuss each of these processes in-turn before providing an overview of the resulting Enron collection sensitivity-aware relevance assessments extension.

\noindent{\textbf{Topic Identification:}} To create our set of sensitivity-aware relevance assessments for the labelled Enron email collection, we first identify a set of topical subjects that reflect the contents of the emails in the collection. Our identified topics are used to form descriptions of information needs that the crowdworkers are asked to provide query formulations for, i.e., example queries that a crowdworker would enter into a search engine to find relevant information to satisfy the information need. We use a topic modelling approach to identify the information needs. When identifying topics to be used as information needs, we are interested in identifying general themes that relate to the topics of discussion that might likely be covered in the contents (i.e., the body) of the emails in the collection. The topics are chosen to be broad enough to be able to reasonably expect that there would be relevant documents in the collection, and not so specific that it would require specialist knowledge to make a judgement of relevance on the subject. We choose to use a topic modelling approach since topic modelling is an efficient way of identifying topics within a collection of documents that is based on statistical modelling of the documents, rather than a blind search through documents attempting to identify topics manually. 

\begin{figure}[tb]
    \centering
    \includegraphics[width=0.45\textwidth]{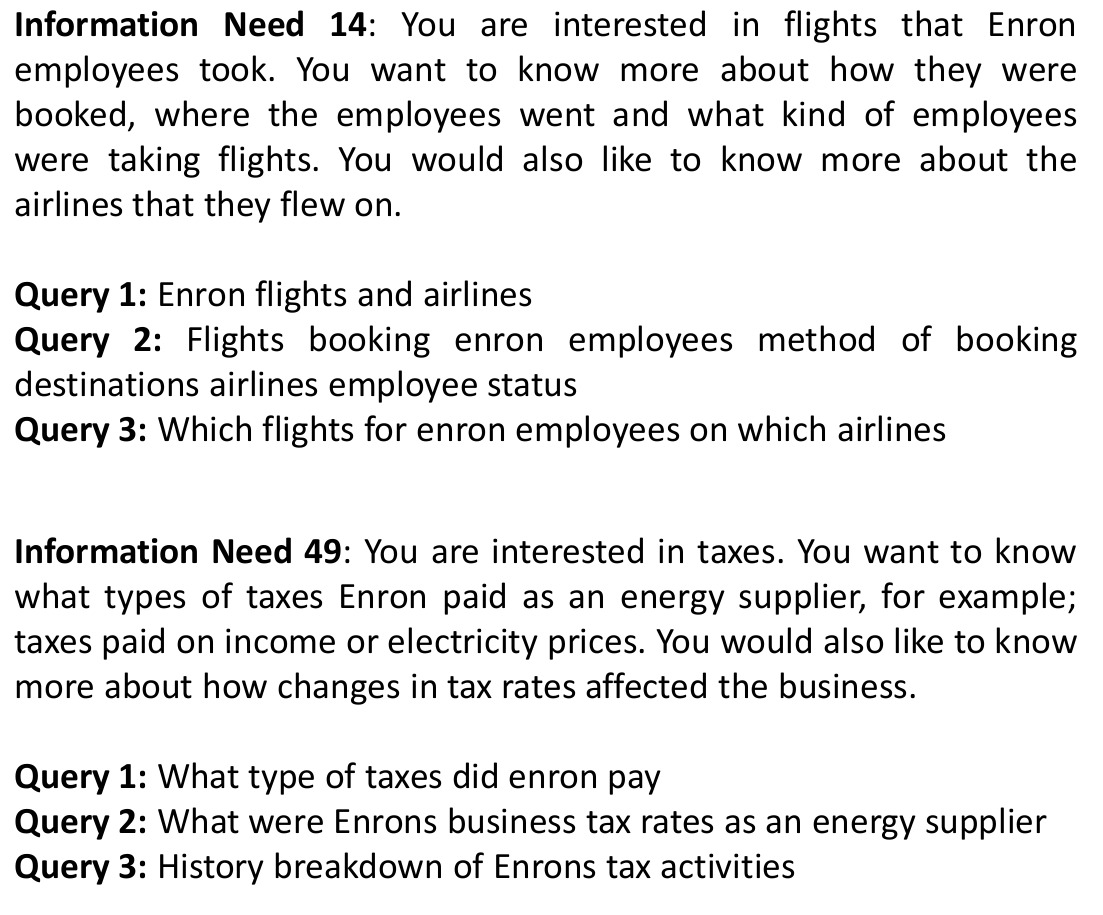}
    \caption{Two example information needs and queries gathered from the crowdsourcing tasks. }
    \label{fig:info_needs}\vspace{-4mm}
\end{figure}

We use the Gensim\footnote{\url{https://radimrehurek.com/gensim/}} implementation of Latent Dirichlet Allocation~\cite{lda} topic modelling to generate fifty topics. As a sanity check to ensure that there are relevant and sensitive documents for each of the identified topics, we select the top 10 terms for each topic and use these as search terms to retrieve documents from the labelled Enron collection. We manually review the top $\sim$ 20 retrieved documents to ensure that at least one relevant document and one sensitive document is retrieved for each of the identified topics. We use PyTerrier~\cite{pyterrier2020ictir} retrieval model to perform this sanity check. 

Subsequently, satisfied that there are at least one relevant and one sensitive emails that are relatively easily retrievable for each of the identified topics, we manually construct short passages of text to serve as descriptions of the information needs that are to be searched for in the collection by the crowdworkers. Figure~\ref{fig:info_needs} presents two illustrative examples of the generated information needs. All of the information needs are of the same structure and are of similar size to the examples in Figure~\ref{fig:info_needs}. The mean number of words in an information need passage is 41.48, with $\sigma^2$ = 27.37 words.

\noindent{\textbf{Crowdsourced Query Formulations:}} In order to collect relevance assessments for pairs of emails and information needs, different query formulations are first needed to generate pools of documents. Query formulations for each topic are collected from crowdworkers from the Prolific crowdwork platform. Ten information needs are shown to each crowdworker and they are asked to provide a query formulation that they would use to get relevant documents to satisfy the information need they are presented with. Ten crowdworkers are recruited for each batch of ten information needs. Attention check questions are used to ensure that the crowdworker is paying attention. These questions are aligned with the guidelines of the Prolific Platform\footnote{\url{https://researcher-help.prolific.co/hc/en-gb/articles/360009223553-Prolific-s-Attention-and-Comprehension-Check-Policy}}. Crowdworkers are presented with a short piece of text which contains the correct - randomised - answer to the question "What is your favourite colour?". The crowdworkers must select the correct option in a multiple choice question. Those who do not answer the attention check correctly are rejected on the Prolific platform. \release{The crowdworkers are paid at a rate of £7/Hour.} Therefore, \release{the output of this crowdworking task is} ten different queries for each of the fifty information needs are collected from crowdworkers, examples of which can be seen in Figure~\ref{fig:info_needs}.

\noindent{\textbf{Pooling Documents for Relevance Assessments:}} Given that there are 1702 documents and 50 information needs, yielding 85,100 information need/query pairs, gathering relevance judgments for every information need, email pair is financially infeasible. Therefore, a select number of documents that are likely to be relevant to the given information need are chosen to be judged~\cite{voorhees_pooling}. This is a pooling approach. Eighteen different retrieval models are used to generate the pools and the top documents in these pools are judged. The eighteen retrieval models were made up of each combination of 3 different queries, three different retrieval models (DPH, BM25, and PL2) and using Bo1 query expansion or not. The three different queries that are used to make the pools are chosen as they provide a diverse set of documents when used with the retrieval models and, after being read by the authors, they are deemed to be well formulated. The top twenty documents in the generated pools are chosen to be judged by crowdworkers. \\

\noindent{\textbf{Crowdsourced Relevance Assessments:}} In order to create a test collection useful for sensitivity-aware search, relevance labels for pairs of information needs and emails are required. Crowdworkers are shown an information need and an email and asked to rate the document as \g{being either} \textit{Highly Relevant}, \textit{Partially Relevant}, or \textit{Not Relevant} to the information need. 

\begin{figure}[tb]
    \centering
    \includegraphics[width=0.4\textwidth]{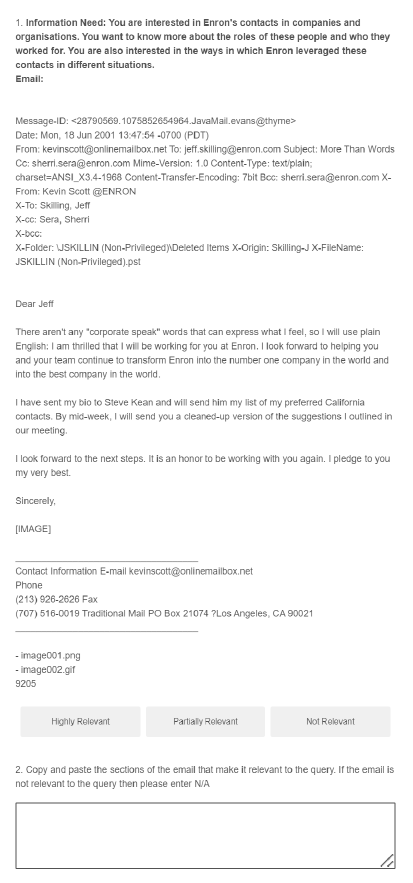}
    \caption{The system used for the relevance assessments crowdworking task}
    \label{fig:crowdworking_system}\vspace{-4mm}
\end{figure}

If a crowdworker judges a document as being relevant, i.e., either \textit{Highly Relevant} or \textit{Not Relevant}, then they are asked to copy and paste the section of the email \g{text} that they thought made the email relevant. This is used as an attention and quality check in addition to the crowdworker being asked to complete an attention check question in the same style as the attention check that is used in the first crowdworking task. \release{The system used for this crowdworking task can be seen in Figure ~\ref{fig:crowdworking_system}. The crowdworkers recruited are paid at a rate of £7/Hour.}

Each information need/document pair is judged by three crowdworkers and a majority vote is used to generate a ground truth label. Since each email-information need pair is judged by three crowdworkers and there are three possible labels, \textit{Highly Relevant}, \textit{Partially Relevant}, and \textit{Not Relevant}, it is possible for each of the labels to be selected by one crowdworker. In practice, this only happened for 134 pairs. In such cases, ties are broken by having one of the authors read the document and make an additional judgement. 

As the relevance labels are being crowdsourced, checks are done to ensure that a number of documents judged as relevant could be returned by baseline retrieval models. In order to ensure that sensitive documents definitely have relevance labels they were also judged by one of the authors for each of the information needs. The crowdsourcing tasks were given ethical approval by the University of Glasgow College of Science and Engineering Ethics Committee. 

\noindent{\textbf{Overview of the Enron Collection Sensitivity-Aware Relevance Assessments Extension:}}

\begin{figure}[t]
    \centering
    \includegraphics[width=0.5\textwidth]{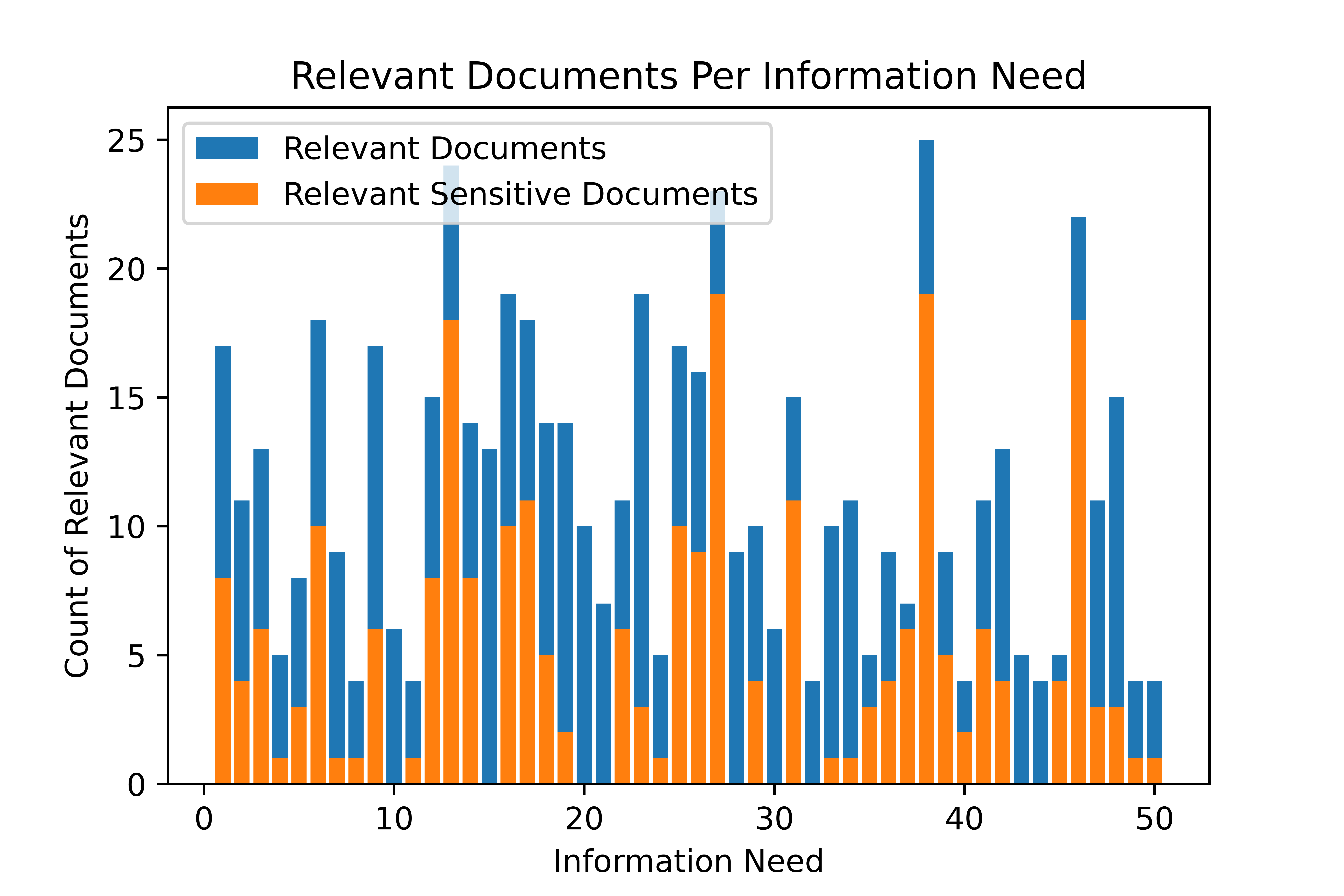}\vspace{-5mm}
    \caption{The number of relevant documents per information need.}
    \label{fig:rel_docs_per_inf_need}\vspace{-4mm}
\end{figure}

The Enron Collection Sensitivity-Aware Relevance Assessments extension consists of fifty different information needs, three different queries for each information need, and 11,471 relevance judgments on a three-point scale of \textit{Highly Relevant}, \textit{Partially Relevant}, and \textit{Not Relevant}. On average, each information need has 11.38 ($\sigma^2$ = 34.73) relevant documents, and on average 6.02 ($\sigma^2$ = 26.72) relevant documents that are sensitive. Figure~\ref{fig:rel_docs_per_inf_need} shows how many relevant and relevant sensitive documents each information need has. Information needs have the same numbering in Figure~\ref{fig:rel_docs_per_inf_need} as in the collection. 

\release{The Enron Collection Sensitivity-Aware Relevance Assessments extension is available at \url{https://github.com/JackMcKechnie/SARA-A-Collection-of-Sensitivity-Aware-Relevance-Assessments}. TSV files that include the queries, relevance assessments, and information needs are made available for download, along with a README file containing information about the collection. \release{The sections in the emails that influenced the crowdworkers decisions on which label to give an email will also be released through the GitHub page.} The files which contain the emails can be downloaded from the UC Berkley website\footnote{\url{https://bailando.berkeley.edu/enron_email.html}}. The Enron Collection Sensitivity-Aware Relevance Assessments extension will also be made available via the popular ir\_datasets library.\footnote{\url{https://ir-datasets.com/}} The dataset is held under an Attribution-NonCommercial 4.0 International (CC BY-NC 4.0) licence which allows for it to be adapted, transformed and built upon.}

\section{Collection Analysis}\label{sec:SASCollectionAnalysis}
The development of sensitivity-aware search engines requires test collections that have classification labels for sensitivity categories and relevance assessments for a set of information needs. Extending the existing Enron test collection with our sensitivity-aware relevance assessments results in such a collection. We now present some preliminary experiments to demonstrate the utility of our extended Enron test collection for developing sensitivity-aware search systems.

Previous work, e.g.,~\citet{oard_sas}, has proposed approaches for developing sensitivity-aware retrieval models that integrate sensitivity predictions directly into the retrieval model. However, when developing a sensitivity-aware search engine, a reasonable baseline approach is to train a sensitivity classifier separately from the retrieval model. The sensitivity predictions can then be applied to the search engine's results to filter out any documents from the results list that are predicted to be sensitive. \citet{oard_sas} refer to this baseline approach as a \textit{post-filtering} approach. In this section, we also deploy post-filtering sensitivity-aware search to illustrate the baseline properties of our extension to the Enron test collection. In the remainder of this section, we first discuss the sensitivity classifiers that we deploy, and their effectiveness, before presenting our sensitivity-aware search experiments.



\begin{table}[tb]
\centering
\caption{Classification scores on the Enron Email Collection.}
\resizebox{0.4\textwidth}{!}{
\begin{tabular}{lcccc} \hline
     & $Precision$       & $Recall$          & $F_1$              & $BAC$             \\ \hline
SVM   & \textbf{0.4856} & 0.5976          & \textbf{0.5358} & 0.7540           \\
LR    & 0.3493          & \textbf{0.6923} & 0.4643          & \textbf{0.7548}
\end{tabular}
}
\label{tab:classification}
\end{table}

\noindent{\textbf{Sensitivity Classification:}} For our experiments, we train two different sensitivity classifiers, namely Support Vector Machine (SVM) and Logistic Regression (LR), to be deployed as two different post-filtering approaches in our sensitivity-aware search experiments. For both of the classifiers, documents (i.e., emails in the labelled Enron collection) are represented by their TF-IDF features. We use a stratified 20\%/80\% train/test split of the existing Enron test collection, resulting in $\sim$ 12\% of the emails in each of the splits being sensitive. The training set is downsampled to account for the unbalanced nature of the sensitivity distribution in the collection. We use scikit-learn\footnote{https://scikit-learn.org/stable/} to train the classifiers (using the default parameters) and to generate the stratified train/test splits. As our classification metrics, we report $Precision$, $Recall$, $F_1$ and $Balanced \text{ } Accuracy$ (BAC). We report BAC since it provides a generalisable measure of the performance of a classifier when the distribution of the class labels is unbalanced, as is the case in our collection. A BAC score of 0.5 denotes random predictions. 

Table~\ref{tab:classification} presents the effectiveness of the sensitivity classifiers. As can be seen from the table, the SVM and LR classifiers correctly classify $sim$60\% and $sim$69\% of the sensitive emails respectively. The SVM classifier makes slightly fewer incorrect predictions than the LR classifier (0.4856 precision SVM vs 0.3493 precision LR), resulting in a higher $F_1$ score for SVM (0.5358 $F_1$ SVM vs 0.4643 $F_1$ LR). The LR classifier achieves the highest BAC score with 0.4578 compared to 0.7540 for the SVM classifier.  

Our deployed classifiers are only intended to illustrate the performance of a baseline sensitivity classification approach in post-filtering sensitivity-aware search. As such, we note that these classification results could be improved upon by deploying more sophisticated classification approaches, e.g.,~\cite{mcdonald_semantic_features}.
 
\noindent{\textbf{Sensitivity-Aware Search:}} In the remainder of this section, we deploy three post-filtering sensitive-aware search approaches to illustrate their performance on our extended Enron test collection. The post-filtering approaches first deploy an off-the-shelf retrieval model to retrieve an initial ranking of documents, before deploying a classifier to predict if each of the documents in the initial ranking are either sensitive or non-sensitive. Documents that are predicted to be sensitive are then removed from the initial ranking to form the final ranked list of results.

For our retrieval model, we deploy the PyTerrier~\cite{pyterrier2020ictir} implementation of BM25 (with default parameters) to retrieve documents from the same 80\% of the documents that is used for evaluating the performance of the classifiers (Table~\ref{tab:classification}). We then deploy each of our trained classifiers, SVM and LR, separately and remove any documents from the ranking that are predicted to be sensitive. This results in two post-filtering approaches, denoted as $BM25\text{-}PostFilter_{SVM}$ and $BM25\text{-}PostFilter_{LR}$ in the remainder of this section. Additionally, we deploy a post-filtering approach that uses the ground truth sensitivity classification labels to make perfect classification predictions and, therefore, removes all of the sensitive documents from the initial ranking, denoted as $BM25\text{-}PostFilter_{Oracle}$. Lastly, we deploy a baseline approach that does not apply any post-filtering to remove sensitive documents, denoted as $BM25\text{-}NoFilter$. When indexing the documents, we remove stopwords and applying Porter stemming using PyTerrier.

We report Precision@10 ($P@10$), Recall@10 ($R@10$), Normalised Discounted Cumulative Gain@10 ($nDCG@10$)~\cite{ndcg} and Cost Sensitive Normalised Discounted Cumulative Gain@10 ($CS\text{-}nDCG@10$) \cite{oard_sas}. $CS\text{-}nDCG@10$ is an extension of Normalised Discounted Cumulative Gain which adds a penalty for returning sensitive documents. In other words, $CS\text{-}nDCG@10$ penalises a retrieval model if it returns a sensitive document in the results list but it allows for the model's recovery if the model returns many relevant (but non-sensitive) documents.



Table~\ref{tab:sas} presents the results of our sensitivity-aware search approaches and demonstrates the utility of our sensitivity-aware relevance assessments extension to the Enron Collection. Our expectation in terms of $CS\text{-}nDCG@10$ is that $BM25\text{-}PostFilter_{Oracle}$ should perform best since it has access to the true classification labels and, therefore, is able to remove all of the sensitive documents from the results list. The trained classifer post-filtering approaches ($BM25\text{-}PostFilter_{LR}$ and $BM25\text{-}PostFilter_{SVM}$) should achieve the next best performances, with the $BM25\text{-}NoFilter$ approach achieving the worst performance in terms of $CS\text{-}nDCG@10$, since it is not able to filter out any sensitive documents. As can be seen from Table~\ref{tab:sas}, our expectation holds with the approaches achieving 0.8623, 0.7280, 0.7236 and 0.7111 respectively. 

Interestingly, we note that the $BM25\text{-}PostFilter_{LR}$ approach results in a higher $nDCG@10$ score compared to the  $BM25\text{-}NoFilter$ approach. This shows that $BM25\text{-}PostFilter_{LR}$ manages to correctly remove non-relevant sensitive documents from the ranking. Overall, these results illustrate the usefulness of the collection for evaluating sensitivity-aware search. Additionally, we note that there is room for improvement in terms of the deployed approaches. These results could be improved upon by deploying more sophisticated sensitivity-aware approaches, such the \textit{$Opt. CS\text{-}nDCG$} approach from~\citet{oard_sas}.





\begin{table}[tb]
\centering
\caption{Sensitivity-aware search scores using SARA}
\resizebox{0.5\textwidth}{!}{%
\begin{tabular}{lcccc} \hline
                          & $P@10$            & $R@10$            & $nDCG@10$         & $CS\text{-}nDCG@10$      \\ \hline
BM25-NoFilter            & \textbf{0.1893} & \textbf{0.2355} & 0.2009          & 0.7111          \\
BM25-PostFilter\_{Oracle} & 0.1601          & 0.1841          & 0.1693          & \textbf{0.8623} \\
BM25-PostFilter\_{SVM}    & 0.1773          & 0.2230           & 0.1945          & 0.7236          \\
BM25-PostFilter\_{LR}     & 0.1773          & 0.2292          & \textbf{0.2044} & 0.7280     
\end{tabular}}
\label{tab:sas}
\end{table}


\section{Discussion}\label{sec:Discussion}
This section provides additional discussion points on the collection. We discuss the previous concerns about the contents of the Enron Email Collection and the size of the collection. 

\subsection{Contents of the Enron Email Collection}
Previous work has had concerns about using the Enron Email Collection for the creation of a dataset for sensitivity-aware search~\cite{oard_collection}. We address these concerns in this section. ~\citet{oard_collection} argued that the ultimate goal of works looking at sensitivity-aware search is to protect sensitive information. Therefore, in their opinion, using crowdworkers to annotate a corpus which contains sensitive information goes against the goal of protecting sensitive information. However, the Enron Email Collection has been publicly available for over a decade and has undergone numerous redaction efforts to remove emails upon request by the authors. Therefore, those ex-Enron employees who felt that they did not want their emails to be read by others have an opportunity over the last 10 years to request the removal of their emails. Additionally, the emails that are shown to the crowdworkers to create our relevance assessments are publicly available and widely used. Consequently, we argue that the use of the Enron Email Collection to build a sensitivity-aware search test collection is justified and provides a valuable resource for the community.

\subsection{Collection Size}
\jmedit{Modern-day information retrieval test collections often contain hundreds of thousands, if not millions, of documents with relevance judgements~\cite{ms_marco, trec_dl_2019}. However, the SARA extension of the Enron Email collection contains only 1702 documents that have been judged for relevance. This is in the nature of the task of sensitivity-aware search - sensitive documents are not often publicly accessible, for the very reason that they are sensitive. Collections that only contain judgements for sensitivity that have been used previously for sensitivity classification are not publicly available and also contain a small number of documents~\cite{mcdonald_semantic_features}.}

\jmedit{In the task of sensitivity-aware search, having small collections that contain one specific genre of sensitivity is desirable. We want to be able to test retrieval systems on a number of test collections, each containing one type of sensitivity, so that we can see how well the system performs on each kind of sensitivity. This is due to the nature of the task and the situations that sensitivity-aware search systems would be deployable to. For example, if we desire to build a sensitivity-aware search engine for searching among the emails of authors donated to an archive, how well that system performs on national security sensitivities is not as important as how it performs on personally identifiable information sensitivities. Consequently, having multiple smaller test collections, that concentrate on one variety of sensitivity is more beneficial, and a more accurate representation of the quality of the system, than a larger collection containing many types of sensitivity.}

\begin{table}[h]
\centering
\caption{Retrieval over all the Enron documents}
\label{tab:all_enron}
\resizebox{0.5\textwidth}{!}{
\begin{tabular}{lcccc}
\hline
\textbf{Name} & \textbf{P@10} & \textbf{R@10} & \textbf{Bpref} & \textbf{nDCG@10} \\ \hline
BM25 & 0.002667 & 0.004611 & 0.126165 & 0.002759 \\
BM25 >> Filter (Oracle) & 0.110667 & 0.106246 & 0.096130 & 0.127596 \\
BM25 >> Filter (Logistic Regression) & 0.000667 & 0.000444 & 0.103885 & 0.000290 \\
BM25 >> Filter (Support Vector Machine) & 0.000667 & 0.000444 & 0.108616 & 0.000290 \\
\end{tabular}
}
\end{table}

\jmedit{If desired, it is possible to retrieve over all (\~500k) documents in the original Enron collection, not just the 1702 that have been judged for relevance and sensitivity. This provides a larger collection to retrieve over, with sparse judgements for relevance. We present experiments demonstrating this, using the same classifiers as defined in Section~\ref{sec:SASCollectionAnalysis}, the results of which are shown in~\ref{tab:all_enron}. In addition to the metrics introduced in Section~\ref{sec:SASCollectionAnalysis}, we also report Binary Preference (BPref)~\cite{bpref} as this measure is more resilient against significant deviations from the completeness assumption of the Cranfield Evaluation Methodology. However, as previously described, within the task of sensitivity-aware search, multiple, smaller test collections containing specific genres of sensitivities is a more relevant measure of the quality of a sensitivity-aware search engine.}

\section{Conclusions}\label{sec:Conclusions}

\release{
In this work, we have identified the need for test collections that are suitable for developing sensitivity-aware search systems. We also presented our sensitivity-aware relevance assessments (SARA) extension to the Enron email collection. LDA topic modelling was performed on the labelled Enron collection to identify topics that are present in the emails. These topics were then used to create a set of 50 information needs. Two crowdsourcing tasks were subsequently carried out. The first task involved sourcing query formulations for the information needs. We crowdsourced three queries per information need (150 queries in total). After carrying out a pooling approach to identify documents that were likely to be relevant to information needs, the second crowdsourcing task was carried out. This involved gathering relevance judgments for information needs. Preliminary experiments using the extended Enron test collection were also performed. Post-filtering sensitivity-aware search approaches were deployed using trained sensitivity classifiers. Our experiments illustrate the usefulness of our sensitivity-aware relevance assessments extension to the Enron email collection for evaluating sensitivity-aware search systems.}


\balance
\bibliographystyle{ACM-Reference-Format}
\bibliography{sas-testcollection}

\end{document}